\documentclass[twocolumn, pre,floatfix]{revtex4}
\usepackage{eurosym}
\usepackage{graphicx,amssymb,amstext,amsmath}
\usepackage{epsfig}
\usepackage[bookmarks=false]{hyperref}
\usepackage{epstopdf}
\usepackage[english]{babel}
\usepackage[autostyle]{csquotes}
\usepackage{lipsum}
\usepackage{subfigure,url}
\usepackage{color,graphics}
\usepackage{braket}
\usepackage{float}
\usepackage{mathtools}
\usepackage[pass,paperwidth=8.5in in,paperheight=11in]{geometry}
\draft
\newcommand*{\addheight}[2][.5ex]{%
  \raisebox{0pt}[\dimexpr\height+(#1)\relax]{#2}%
}
\begin{document}

\title{QUANTUM MONA LISA CAT} 

\author{Rashid Ahmad, Sumaira Nawaz}

\affiliation{Department of Physics, Kohat University of Science and Technology, Kohat 26000, Khyber-Pakhtunkhwa, Pakistan}

\date{\today}
\begin{abstract}
Schr\"{o}dinger's Cat was proposed by Erwin Schr\"{o}dinger, the infamous thought experiment in which a cat in a box was both alive and dead simultaneously illustrating a quantum phenomenon known as superposition. In 2013, Yakir Aharonov and his co-authors conceived of an experiment suggesting that a particle can be separated from its property. They called the effect a "Quantum Cheshire Cat" that has been experimentally verified in the succeeding year. The name Quantum Cheshire Cat is inspired from a fanciful character of the Cheshire Cat in "\textit{Alice's Adventures in Wonderland"} a novel written by Lewis Carroll where the grin of cat is found without a cat. An important question arises here. Once the grin of the Cheshire Cat is separated, is there any correlation still left between the grin and the cat? To answer the question we propose a thought experiment in which Quantum Cheshire Cat is also a Schr\"{o}dinger's Cat existing in superposition of  happy(smiling) and sad(frowning) states. We name this cat as a "Quantum Mona Lisa Cat" for the reason that historically it is presumed that Mona Lisa's portrait contains both characteristics of happy(smiling) and sad(frowning) and either is observed depending upon the mood of the observer. We show that property separated from particle behave as  "Quantum Mona Lisa Cat".
\end{abstract}
\keywords{}

\maketitle 


\section{Introduction}
Superposition Principle is the core principle of quantum mechanics where a particle can be in a superposition of many states\cite{a}. After measurement it collapses to one of the basis states that form the superposition, thus destroying the original configuration. To explain this principle and to make it convenient for the scientific community Erwin Schr\"{o}dinger suggested in $1935$ a thought experiment where a living Cat is put into a steel chamber alongside a sledge, a vial of hydrocyanic corrosive and a limited quantity of radioactive substance. In the event that even a solitary atom of the radioactive substance rots during the trial, a hand-off component will trip the mallet, which will, thus, break the vial of noxious gas and cause the Cat to kick the bucket. Until the container is open Cat exists in the two conditions of dead what's more, alive. When one opens the box, the wavefunction collapses to one of its two possible states.

\par
Rather than the standard von Neumann or strong-Measurements, which gives data about the eigenvalues of a dynamical operators, Weak-Measurements empower us to acquire data about little prompted stage changes. Weak measurements can uncover some data about the amplitudes of a quantum state without crumbling the state into eigenvectors. This is finished by a feeble coupling between the measurement device and the system.
\par
Consider a quantum framework in pure-state $| \psi \rangle$, if extent of an observable A is taken, the result is a random eigenvalue $\alpha_{i}$ of A. Estimation of A on states, presents that the estimation results are probabilistic, the likelihood of result $\alpha_{i}$ in some random run being $|\langle \alpha_{i}| \Psi \rangle|^{2}$, where $|\alpha_{i}\rangle$ is the eigenstate, consistent to the eigenvalue $\alpha_{i}$. Along these lines in quantum mechanics, one measures the expectation value $\langle A \rangle$ that is the collective average of results. What's more, in a path, the condition of the system collapses to $|\alpha_{i}\rangle$. Thus, there is no broad accord among physicists concerning whether the estimation of a discernible uncovers a property of the framework or is an ancient rarity of the estimation procedure itself. This is achieved by weakly measuring the observable A, making insignificant unsettling influence the state. We quickly summarize the procedure of weak measurements underneath.
\par
Suppose a quantum framework is pre-selected in the state $| \Psi_{i} \rangle$. Consider that an observable A is weakly-measured by presenting a minor connection between the quantum framework and a proper measuring meter. For all successful post-selections of the state $| \Psi_{f} \rangle$, the meter evaluations consistent to the weak-measurements of A are taken into concern whereas the others are cast-off. The alteration in the meter evaluations, typical, for all such post-selected systems, is $A_{\omega}$ which is the weak-value of A \cite{1}. Mathematically the weak-value of A is defined as;

\begin{align}
\langle A_{\omega}\rangle = \frac{\langle \Psi_{f} |A| \Psi_{i} \rangle}{\langle \Psi_{f}|\Psi_{i} \rangle}
\end{align}
Although interpreted as a value of an observable, the weak value can lie outside the eigenvalue spectrum \cite{1,2,3} and can even be complex \cite{4} with the imaginary part being related to the shift in the momentum of the pointer. Weak values have already been experimentally observed  \cite{5,6,7,8,9,10,11,12}.
\par
It is generally accepted that the properties of a physical framework can't be isolated from the structure itself. This image of the plausibility of physical-systems, yet, does not remain enduring in the territory of quantum-mechanics. A recent study, known as the Quantum Chesire Cat \cite{13}, shows that property, for instance, the polarization of a photon can be separated from the photon. In light of an altered form of a Mach-Zehnder interferometer, the Quantum Cheshire Cat shows that a photon can be separated from its circular polarization and can lead to travel freely along the two arms. In such experiments, the technique used is Weak measurements and it is of most extreme significance to deliberately pick pre-and post-selected ensemble, which means we need to set up our particles in initial state $|\Psi_{i}\rangle$ and when they leave our experimental arrangement we perform post-determination with the goal that we get final state $|\psi_{f}\rangle$. Just on the off chance that states $|\Psi_{i}\rangle$ and $|\Psi_{f}\rangle$ are picked successfully, at that point we can watch Quantum Cheshire Cat.
\par
Quantum Cheshire Cat lightened up more knowledge of quantum frameworks and pulled in a ton of discussions \cite{14,15,16}. It relates not entirely to photons and their polarizations yet can, on a basic level, be seen with any quantum framework and its property. Trial confirmations of the wonder with a neutron as the Cat and its spin as the smile have been reported \cite{17,18}. The impact is used to understand the three-box Paradox \cite{19} and is contemplated within the sight of decoherence \cite{20}. As of late, a convention has been created utilizing which the separated grin of Quantum Cheshire Cat has been transported between two spatially isolated gatherings without Cat \cite{21}.
\par
The Quantum Cheshire Cat has been observed experimentally using neutron interferometry \cite{8} as well as photon interferometry \cite{9,11,12}.  A refined version of the original proposal has been suggested in \cite{23} that decouples all the components of the polarization from the photon. Another proposal that deals with the separation of two degrees of freedom belonging to the same photon can be found in \cite{24}. Some of the recent works in this area include the teleportation of the decoupled circular polarization without the photon \cite{25} and exchanging the decoupled circular polarizations of two Quantum Cheshire photons \cite{26}.  Further discussions on the topic can be found in \cite{15,29,James}.
\par
Now Mona Lisa,likely the most acclaimed portrait in the history was painted by Leonardo Da Vinci between $1503$
and $1514$ shown in Fig. \ref{fig:mona} and is on perpetual presentation at the Louver Museum in Paris. One of the unique things that lead to the popularity of Mona Lisa's portrait is her puzzling articulation, which seems both alluring and aloof and has given portrait universal fame. The question was: Did Leonardo Da Vinci paint Mona Lisa with a smile or a frown? Art lovers have long debated over the famous painting, where Mona Lisa appears to have a changing expression but a new study from experts at the University of California could settle the arguments once and for all. According to the researchers, a person's perception of Mona Lisa's expression depends on their disposition at the time i.e. contextuality of the observer. On the off chance that one sees the Mona Lisa after having a shouting battle, one is going to see it in an unexpected way, but in case one is having the cheerful  great time,  the cryptic grin will appear \cite{30}. Here, we give a new dimension to the portrait by saying it is in the superposition of happy and sad states and it collapses to either of the states depending upon the mood of the observer.
\begin{center}
\noindent
\begin{figure}
\begin{minipage}{1.0\columnwidth}

      \addheight{\includegraphics[width=80mm, height=50mm]{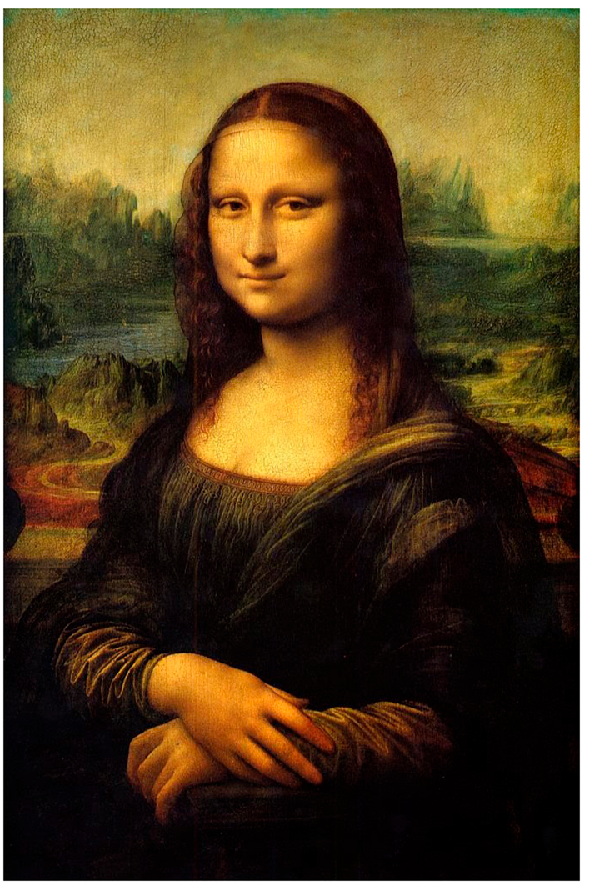}}

\caption{Portrait of Mona Lisa}
\label{fig:mona}
\end{minipage}
\end{figure}
\end{center}

\par
We consider helicity as a parameter for determining whether particle and its property have some correlation even after they are physically separated. Let's assume that particle momentum plays the role of mood of the observer and property is the smile or frown of Mona Lisa portrait. So when observer mood changes i.e. the momentum gets changed so as the expression on the portrait changes i.e. the orientation of the property changes. We use this analogy to call the system a Quantum Mona Lisa Cat.
\section{Quantum Mona Lisa Cat}
When a particle is separated from its property it's important to know that whether there is still some correlation left between property and particle. The helicity of the particle can be used to determine the presence of any correlation between them both for massive and massless particles.  We here consider the case of spin separated from the electron, however, it can be generalized to other properties and particles.
\par
Helicity is defined as the projection of a spin vector \textbf{s} in the direction of its momentum vector \textbf{p}
\begin{align}
H\equiv\frac{\textbf{s}.\textbf{p}}{|\textbf{s}.\textbf{p}|}.
\end{align}

We use the technique of weak-measurements to find the sign of helicity by taking pre-selected state as
\begin{align}
| \Psi_{1} \rangle = \frac{1}{\sqrt{2}}(L_{p}|1\rangle + (\sigma_{\uparrow}+\sigma_{\downarrow})|2\rangle).
\end{align}

Passing this through the modified Mach-Zehnder interferometer consisting of Beam Splitters (BS), Magnetic Field (B), Phase Shifter (PS), Spin Turner (ST) and Analyzer (A) shown in Fig. \ref{fig:PH}.

\begin{center}
\noindent
\begin{figure}
\begin{minipage}{1.0\columnwidth}

      \addheight{\includegraphics[width=85mm]{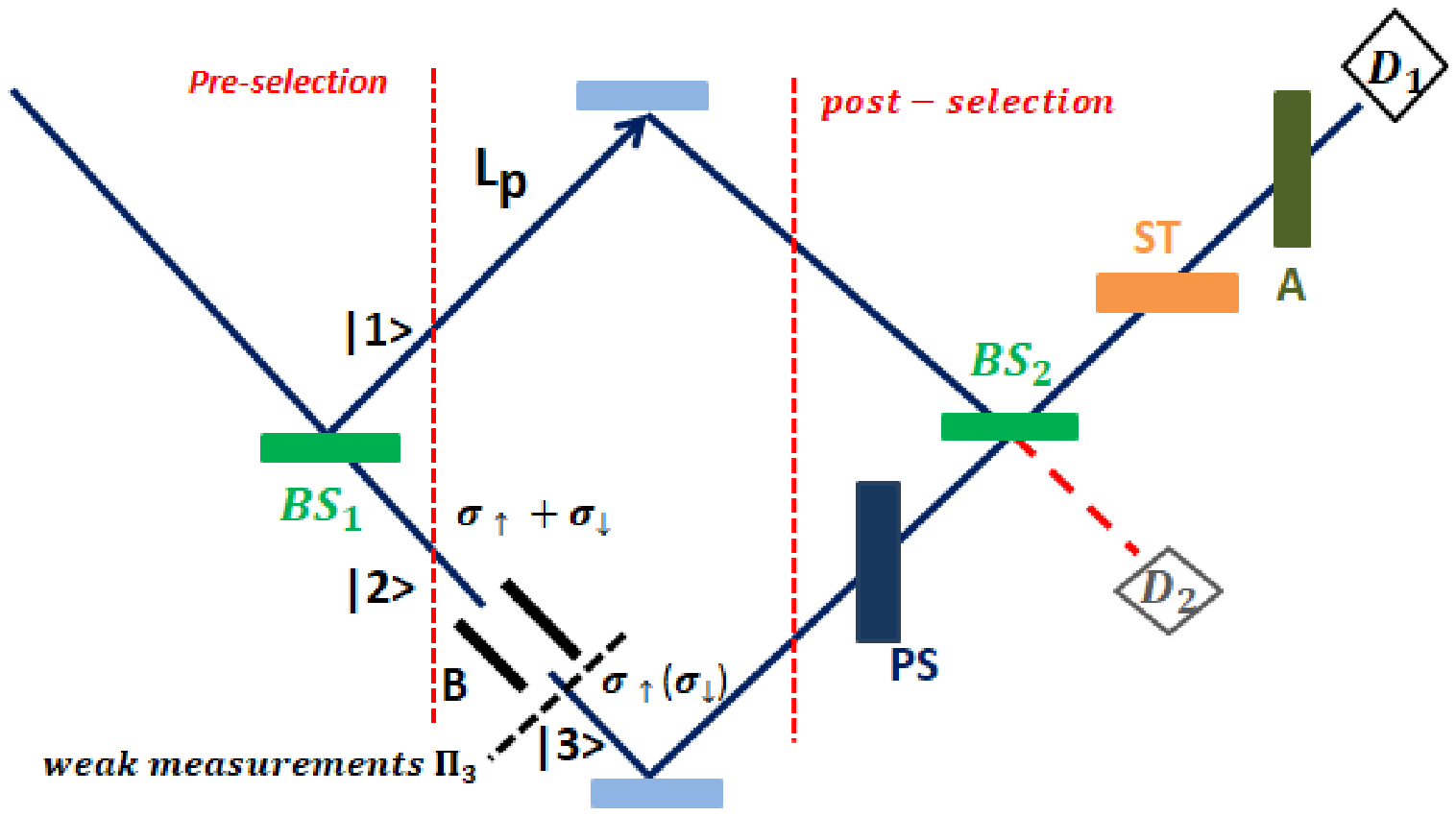}}

\caption{Electron having momentum \textbf{$L_{p}$} travels along path $|1\rangle$, and spin travels along path $|2\rangle$. Spin is in superposition of spin-up \textbf{$(\sigma_{\uparrow})$} and spin-down \textbf{$(\sigma_{\downarrow})$} states. Spin is passed through magnetic field \textbf{B} and collapses into spin-up \textbf{$(\sigma_{\uparrow})$}\big(spin-down \textbf{$(\sigma_{\downarrow})$}) state which  travels along the path  $|3\rangle$ . Various measuring devices are inserted into left and right arms of the interferometer, between the pre- and post-selection.}
\label{fig:PH}
\end{minipage}
\end{figure}
\end{center}

We have the post-selected state
\begin{align}
| \phi_{1} \rangle = \frac{1}{\sqrt{2}}(L_{p+}|1\rangle +\sigma_{\uparrow}|3\rangle),
\end{align}

for measuring weak-values let us define Projection operators
\begin{equation}
\Pi^{(1)}_{p+}= |1\rangle\langle 1|,
\end{equation}
\begin{equation}
\sigma^{(3)}_{\uparrow}= \Pi_{3}\sigma_{\uparrow},
\end{equation}
where
\begin{equation}
\Pi_{2}= |2\rangle\langle 2|,
\end{equation}
similarly
\begin{equation*}
\sigma^{(4)}_{\downarrow}= \Pi_{4}\sigma_{\downarrow}.
\end{equation*}

The measuring device is modified version of Mach-Zehnder interferometer consisting of beam splitters $(BS_{1})$ and $(BS_{2})$, phase shifter (PS), a Spin Turner (ST), Analyzer (A) and two electron indicators. Given these choices, if the state preceding the PS is $|\phi_{1} \rangle$, at that point $D_{1}$ will click with assurance. An electron in any state orthogonal to $|\phi_{1} \rangle$ will wind up at detector $D_{2}$. We will concentrate just on those cases wherein detector $D_{1}$ clicks, as investigated for neutrons in \cite{17}. Let us first inquire what direction the electron will go inside the interferometer. We will show that, given the pre- and post-selection, with certainty the electron with momentum $L_{p}$ went through the arm $|1\rangle$. Suppose that we check the location of the electron by inserting electron detectors into the arms of the interferometer. We assume that the detectors be non-demolition detectors in the sense that they do not absorb the electron and do not alter its spin. In mathematical terms, these detectors measure the projection operators given by above equations. Suppose first that we insert one such detector into the arm $|2\rangle$. Is it conceivable to find the electron there? No, it isn't. On the off chance that we find an electron there, at that point the state after this estimation will be $| \phi^{'}_{1} \rangle = L_{p+}|2\rangle$ , which is orthogonal to the post-selected state $| \phi_{1} \rangle = \frac{1}{\sqrt{2}}(L_{p+}|1\rangle +\sigma_{\uparrow}|3\rangle)$. Hence the post-selection could not have succeeded in this case (i.e. detector $D_{1}$ could not have clicked). Thus the non demolition measurement in the arm $|2\rangle$ never finds the electron there, indicating that the electron must have gone through the arm $|1\rangle$.On the off chance that rather we play out a non-destruction estimation in the arm $|1\rangle$, given the post-selection, it will always indicate that the electron is there.We can even perform non-destruction estimations in the two arms all the while, furthermore, they will consistently show that the electron was in the arm $|1\rangle$.

To find out the sign of helicity, measuring weak-value at path $|3\rangle$ will result
\begin{align}
\langle\Pi_3\rangle_{\omega} =\frac{\langle \Psi_{1} |\Pi_3| \Phi_{1} \rangle}{\langle \Psi_{1}|\Phi_{1}\rangle}=1.
\end{align}
The detector $D_{1}$ will click with certainty as the weak-value is $1$. This result reveals that the helicity is positive.

\par
If the weak value $\langle\Pi_3\rangle_{\omega}$ is $0$, detector $D_{1}$ will not click and the helicity will be considered as negative, as shown in Fig. \ref{fig:PH}. After determining the nature of helicity, we will now find that whether by reversing the direction of momentum \textbf{p}, the orientation of spin \textbf{s} also reverses as shown in Fig. \ref{fig:CONH}. Here, we are not considering the inversion of helicity because of the boost. We are using the tool of weak-measurements with pre and post-selected states as
\begin{align}
| \psi_{2} \rangle = \frac{1}{\sqrt{2}}|L_{-p}\rangle|5\rangle +|\sigma_{\uparrow}\rangle|3\rangle,
\end{align}

and the post-selected state
\begin{align}
| \phi_{2}\rangle = \frac{1}{\sqrt{2}}|L_{-p}\rangle|5\rangle +|\sigma_{\downarrow}\rangle|4\rangle.
\end{align}
In other words, we are going to perform our work in such a way that the post-selected state is surely achieved. This is the helicity preserving case.

\begin{center}
\noindent
\begin{figure}
\begin{minipage}{1.0\columnwidth}

      \addheight{\includegraphics[width=85mm]{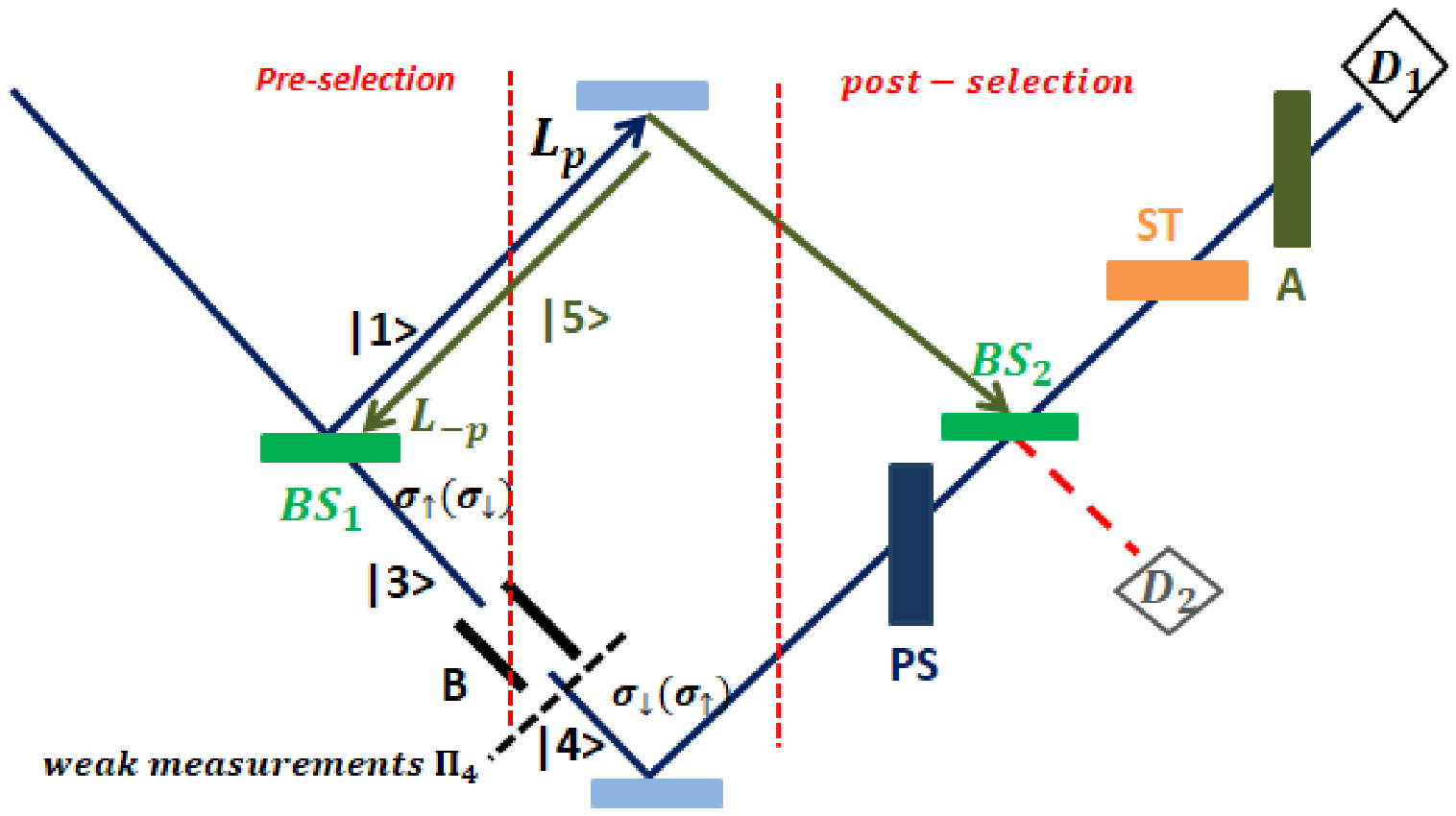}}

\caption{Direction of momentum \textbf{$L_{p}$} is reversed to \textbf{-p} along the path $|5\rangle$. The orientation of spin changes from spin-up \textbf{$(\sigma_{\uparrow})$}\big(spin-down \textbf{$(\sigma_{\downarrow})$}\big) (path $|3\rangle$ ) to spin-down \textbf{$(\sigma_{\downarrow})$} \big(spin-up \textbf{$(\sigma_{\uparrow})$}\big) travelling along the path $|4\rangle$.}
\label{fig:CONH}
\end{minipage}
\end{figure}
\end{center}

Measuring weak-values at path $|4\rangle$
\begin{align}
\langle\Pi_{4}\rangle_{\omega} =\frac{\langle \Psi_{2} |\Pi_{4}| \Phi_{2} \rangle}{\langle \Psi_{2}|\Phi_{2}\rangle}=1.
\end{align}

Weak-measurement at path $|4\rangle$ gives a weak-value $1$, confirming the statement that helicity is conserved i.e. by reversing direction of  momentum \textbf{p} the spin \textbf{s} also reverses. This gives the idea that there is still some sort of quantum correlation between the spin and particle although, they are separated as shown in Fig. \ref{fig:CONH}. Helicity reversing case may be obtained when the pre and post-selected states are
\begin{align}
| \psi _{3}\rangle = \frac{(L_{p}|5\rangle +\sigma_{\uparrow}|3\rangle)}{\sqrt{2}},
\end{align}

and the post-selected state
\begin{align}
| \phi_{3} \rangle = \frac{(L_{-p}|5\rangle +\sigma_{\uparrow}|4\rangle)}{\sqrt{2}}.
\end{align}

Measuring weak-values at path $|4\rangle$
\begin{align}
\langle\Pi_4{\omega}\rangle = \frac{\langle \Psi_{3} |\Pi_4| \Phi_{3} \rangle}{\langle \Psi_{3}|\Phi_{3}\rangle}=1.
\end{align}

By reversing the direction of momentum the spin does not change its orientation meaning that quantum property like the spin, is completely separated from the particle.

\section{Conclusion}
The idea of Quantum Mona Lisa Cat depicts that property separated from the particle is still correlated to the particle.  Separated-spin but entangled with a particle can have applications for fundamental quantum mechanics and in fields like quantum computing, quantum walks, and quantum optics. Our thought experiment demonstrates that dislocated spin from its particle can still be influenced by the change in some other parameter located with the particle. Such entanglement can be extended to correlate other properties as well for example correlation of charge and spin will be worth of further investigations.  The case of more than two particles entangled initially might be interesting situation to be investigated for the Quantum Mona Lisa Cat scenario.

\end{document}